\documentclass[twocolumn,showpacs,prb]{revtex4}

\usepackage{epsfig,graphicx,hhline}

\usepackage{dcolumn}
\usepackage{bm}

\begin{document}

\title{Electronic structure of silver-deficient hexagonal AgB$_2$}

\author{I.R. Shein$^*$, N.I.Medvedeva and A.L. Ivanovskii}

\affiliation {Institute of Solid State Chemistry, Ural Branch of
the Russian Academy of Sciences, 620219, Ekaterinburg, Russia}

\begin{abstract}
Electronic structure and cohesive properties of metastable
hexagonal AgB$_{2}$ and silver-deficient borides
Ag$_{0.875}$B$_{2}$ and Ag$_{0.750}$B$_{2}$ were investigated by
means of the projected augmented wave method in the framework of
the density functional theory (VASP package). We found that the
density of states at the Fermi level for nonstoichiometric
diborides is almost constant within a range of vacancy content up
to 25\%. The formation energy of metal vacancies in silver
diboride is the least among all 4d metal diborides,  i.e. for
AgB$_{2}$  is possible to expect the wide homogeneity region.\\

$^*$ E-mail: shein@ihim.uran.ru
\end{abstract}


\maketitle After the discovery of superconductivity with T$_c\sim$
39 K\cite{Akimitsu} in layered MgB$_2$, there have been a lot of
theoretical searches for the potential superconductors with high
T$_c$ among other diborides MB$_c$ (M = Na, Be, Ca, noble and
transition
metals)\cite{Vajeeston,Suzuki,Shein1,Medvedeva,Mehl,Kwon,Yang,Oguchi,Shein2,Profeta},
see also review \cite{Ivanovskii1}. Among them the silver
diboride (AgB$_2$) was predicted to be promising system having
the larger density of B2p $\sigma$-like states near the Fermi
level\cite{Shein1} and electron-phonon coupling constant than
MgB$_2$ and hence higher T$_c$ ($\sim$ 60 $\div$ 70
K)\cite{Kwon,Yang}.\\
However the attempts to synthesize AgB$_2$ lead to inconsistent
results. For the first time the synthesis of AgB$_2$ phase (space
group P6/mmm, lattice parameters a = 3.00 and c = 3.24 \AA) has
been declared thirty years ago\cite{Gurin}. Further the different
experimental routes were used\cite{Sinder} to obtain AgB$_2$
samples as a bulk and thin films; however all synthesized AgB$_2$
samples were unstable.\\
Quite recently the successful synthesis of silver boride thin
films with nominal composition AgB$_2$ was performed by a pulsed
laser deposition method\cite{Tomita} and sharp superconducting
transition was observed on the resistivity with the onset of
superconductivity at 7.4 K and the zero resistance near 6.7 K.
Thus the measurements\cite{Tomita} confirmed the
superconductivity in the silver boride system. However the
observed T$_c$ was significantly lower than the theoretically
predicted value\cite{Kwon,Yang} and comparable with T$_c$ for
some $\textit{d}$ metal diborides: ZrB$_2$ (5.5 K), TaB$_2$ (9.5
K) and NbB$_2$ (5.2K)\cite{Ivanovskii1}.\\
This discrepancy may be caused by the inhomogeneous structure of
the AgB$_2$ films, in particular by nonstoichiometry of samples.
Recently was established\cite{Yamamoto,Takeya,Takeya1}, that some
d metal diborides at nonequilibrium conditions may contain a
significant amount of metal vacancies (for Nb$_{1-x}$B$_2$ and
Ta$_{1-x}$B$_2$ up to x $\sim$ 0.48). The band structure
calculations\cite{Shein3} showed that the metal vacancies are
most likely to appear in diborides with rather weak M-B covalent
bonding. It may be expected the presence of significant
concentration of Ag vacancies in metastable AgB$_2$.\\

\begin{table}
\begin{center}
\caption{Lattice parameters $\textit{a}$ and $\textit{c}$ (\AA),
cell volume V$_0$ (\AA$^3$), bulk modulus B$_0$ (GPa), its
pressure derivative B'$_0$, intra- $\alpha_a$ and interlayer
$\alpha_c$ (TPa$^{-1}$) linear compressibility for AgB$_2$:
experiment and theory.}

\begin{tabular}{|c|c|c|c|c|c|}
\hline
& \multicolumn{2}{c|}{ }&\multicolumn{3}{c|}{ }\\
&\multicolumn{2}{c|}{Experiment}&\multicolumn{3}{c|}{Theory\footnote{FLAPW \cite{Kwon} and LCAO SCF Hartree-Fock\cite{Parvin} calculations}}\\
Parameters& \multicolumn{2}{c|}{ }&\multicolumn{3}{c|}{ }\\
\cline{2-6} & & & & & \\ & see\cite{Gurin} & & see\cite{Kwon}& see\cite{Parvin} & Our data\\
& & & & & \\
\hline
& & & & & \\
$\textit{a}$ & 3.000 & 3.040\cite{Cheng}\footnote{extrapolated from the Vegard relationship for Ag-doped MgB$_2$} & 2.980 & 3.000 & 3.024\\
$\textit{c}$ & 3.024 & 3.180\cite{Sinder} & 3.920 & 3.617 & 4.085\\
$\textit{c/a}$ & 1.0800 & - & 1.3154 & 1.2051 & 1.3414\\
& & & & & \\
\hline
& & & & & \\
V$_0$ & 25.253 & - & 30.147 & 28.188 & 32.361\\
& & & & & \\
\hline
& & & & & \\
B$_0$ & - & - & - & 239  & 142 \\
B'$_0$& - & - & - & 3.20 & 4.00\\
& & & & & \\
\hline
& & & & & \\
$\alpha_a$\footnote{The isostatic pressure effect on the lattice
constants generally expressed as: a = a$_0$(1 + $\alpha_a$P) and
c = c$_0$(1 + $\alpha_c$P), where P is the pressure in GPa,
$\alpha_a$ and $\alpha_c$ are axial compressibilities}&-&-&-&-0.93&-1.35\\
$\alpha_c$&-&-&-&-2.10&-3.06\\
& & & & & \\
\hline
\end{tabular}
\end{center}
\end{table}

In this paper, we concentrate on the effects of Ag vacancies on
the electronic and cohesive properties of silver diboride. For
this purpose, the electronic structure of hexagonal AgB$_2$, as
well as Ag-deficient Ag$_{1-x}$B$_2$ (x = 0.875 and 0.750) are
investigated theoretically using the projected augmented wave
(PAW) method in the framework of the density functional theory
(the Vienna ab initio simulation package -
VASP\cite{Kresse1,Kresse2,Kresse3,Kresse4}) with generalized
gradient approximation (GGA) for exchange-correlation
potential\cite{Perdew96}.\\
The silver diboride have the hexagonal crystal structure (space
group P6/mmm) composed of layers of trigonal prisms of Ag atoms
in the center of boron atoms, which form the planar graphite-like
networks. Both complete and nonstoichiometric AgB$_2$ phases were
simulated by the 24-atoms supercell (Ag$_8$B$_{16}$). The removing
of one silver atom (Ag$_7$V$^{Ag}$B$_{16}$, where V$^{Ag}$ is the
silver vacancy) describes the Ag$_{0.875}$B$_{2}$ composition. For
the Ag$_{0.750}$B$_2$ composition (supercell
Ag$_6$V$_2^{Ag}$B$_{16}$) some possible distributions of Ag
vacancies are checked. For this purpose two vacancies (in cell)
were located at the nearest positions in the Ag sheets, i.e. the
alternation of complete and V$^{Ag}$ - containing sheets along
c-axis is described (case I). Secondly, two vacancies were placed
in the neighbor sheets, and the "uniform" distribution of
vacancies in a crystal was modeled (case II), Fig. 1. For all
systems the structural parameters $\textit{a}$ and $\textit{c}$ have been optimized.\\
Let's discuss the data concerning an ideal AgB$_2$ phase. Table 1
shows the optimized lattice parameters of AgB$_2$ as well as the
zero pressure bulk modulus B$_0$ and its pressure derivative
(B'$_0$ = dB$_0$/dP) in comparison with some previous
experimental and theoretical results. There is a significant
difference between the contraction rates of intra- and
inter-planar periodicity: the compressibility in AgB$_2$ is
strongly anisotropic and the structure is more compressible in
the z direction (c=c$_0$ (1- 0.00360 P)) than in the xy plane (a
= a$_0$(1 - 0.00106 P))(P in GPa). Thus, the interlayer linear
compressibility exceeds the intra-layer linear compressibility
about 2.3 times.\\
The calculated heat of formation ($\Delta$H, defined as a
difference in the total energies of AgB$_2$ with reference to the
E$_{tot}$ of the constituent elements in their stable
modifications: fcc-Ag and rhombohedral boron ($\alpha$-B$_{12}$))
has a small, but negative value (-0.98 eV/f.u.) testifying the
possibility of AgB$_2$ synthesis. For comparison, $\Delta$H for
stable refractory 4d metal diborides ZrB$_2$ and NbB$_2$ were
found to be -4.76 and -3.68 eV/f.u., respectively (FLMTO
calculations\cite{Shein3}). The cohesive energy (E$_{coh}$ $\sim$
14.1 eV/f.u.) is also minimal for AgB$_2$ in comparison with
other 4d metal diborides: 23.4, 24.8 and 19.3 eV/f.u. for
ZrB$_2$, NbB$_2$ and YB$_2$\cite{Shein3}. The metastable nature
of AgB$_2$ is determined by the very weak inter-layer Ag-B and
intra-layer Ag-Ag interactions. The bonding picture may be
examined using the difference electron densities ($\Delta\rho$),
where neutral atomic charge $\rho^{at}$ densities are subtracted
from the crystalline density $\rho^{cryst}$, Fig.2. Negative
$\Delta\rho$ around Ag and positive ones around boron indicate a
charge transfer from Ag to B. The $\Delta\rho$ map shows a
strongly covalent B-B bonding in the hexagonal boron sheets. On
the contrary, the partial ionic type of the inter-plane Ag-B
bonding occurs. The in-plane Ag-Ag bonds are insignificant: (i)
the near-spherical symmetry of silver $\Delta\rho$ contours
confirm the absence of Ag-Ag covalency; (ii) the ionic interaction
between silver ions is repulsive and (iii) the Ag-Ag distance in
diboride (3.004 \AA) is about 4\% higher than in fcc Ag (2.889
\AA) i.e. the decrease in metallic-like bonding also occurs.\\
The electronic band structure of AgB$_2$ obtained in our
calculations coincides with the previous
results\cite{Shein1,Kwon}. The B2p $\sigma$-like (in-plane) bands
have a small dispersion along $\Gamma$-A-L in the BZ the and form
the hole Fermi surfaces\cite{Shein1}. The most attractive feature
of AgB$_2$ is that the B2p $\sigma$-like bands are more flat than
in MgB$_2$, yielding the higher density of states at the Fermi
level (N(E$_F$)). We found that N(E$_F$) = 0.90 states/eVcell,
it is about 20\% larger that in MgB$_2$ (N(E$_F$)= 0.72
states/eVcell). Calculated densities of states for AgB$_2$,
Ag$_{0.875}$B$_2$ and Ag$_{0.750}$B$_2$ are shown in Fig. 3. The
occupied Ag 4d bands are located mainly in the region of -5.8
$\div$ -3.3 eV. The upper part of conduction band is composed by
comparable contributions of Ag 4d  and antibonding B 2p states.\\
Let us consider the effect of silver vacancies on the structural,
cohesion and electronic properties of AgB$_2$. The introducing of
V$^{Ag}$ leads to the reduction of both lattice constants; however
for x = 0.125 the c/a ratio decreases, whereas for x = 0.250 this
parameter grows and it appears larger, than that for complete
AgB$_2$, Table 2. The inter-planar compression is more pronounced
for the "uniform" V$^{Ag}$ distribution (model II).\\
The energies of vacancy formation (E$_{vf}$) as well as E$_{coh}$
and $\Delta$H for Ag$_{1-x}$B$_{2}$ phases are summarized in
Table 2. These parameters are calculated as:

\begin{equation}
E_{vf}=E_{tot}^{AgB_{2}}-E_{tot}^{Ag_{1-x}B_{2}}-xE_{tot}^{Ag},
\end{equation}

\begin{equation}
E_{coh}^{Ag_{1-x}B_{2}}=E_{tot}^{Ag_{1-x}B_{2}}-[(1-x)E_{at}^{Ag}+2E_{at}^{B}],
\end{equation}

\begin{equation}
\Delta
H^{Ag_{1-x}B_{2}}=E_{tot}^{Ag_{1-x}B_{2}}-[(1-x)E_{tot}^{Ag}+2E_{tot}^{B}],
\end{equation}\\

where E$^{Ag}_{at}$, E$^B_{at}$ are the total energies of free
silver and boron atoms, E$_{tot}^{Ag}$, E$_{tot}^B$ are the total
energies of fcc Ag metal and $\alpha$-boron; and
E$_{tot}^{AgB_2}$, E$_{tot}^{Ag_{1-x}B_2}$ are the total energies
(per formula units) of AgB$_2$ and Ag$_{1-x}$B$_2$,
respectively.\\

\begin{table}
\begin{center}
\caption{Lattice parameters (\AA), cohesive energies (E$_{coh}$),
heat of formation ($\Delta$ H) and vacancy formation energies
(E$_{vf}$) in eV/f.u. for complete and silver-deficient Ag
borides.}
\begin{tabular}{|c|c|c|c|c|c|c|}
\hline
& & & & & &\\
Diborides & $\textit{a}$ & $\textit{c}$ & $\textit{c/a}$ &
E$_{coh}$ & $\Delta$
H & E$_{vf}$\\
& & & & & &\\
\hline
& & & & & &\\
AgB$_2$              & 3.024 & 4.085 & 1.3414 & 14.06 & -0.98 & -\\
& & & & & &\\
Ag$_{0.875}$B$_2$    & 3.008 & 4.020 & 1.3364 & 13.71 & -0.64 & 0.34\\
& & & & & &\\
Ag$_{0.75}$B$_2$(I)& 2.942 & 3.960 & 1.3460 & 13.32 & -0.26 & 0.71 \\
& & & & & &\\
Ag$_{0.75}$B$_2$(II)& 2.950 & 3.950 & 1.3551 & 13.35 & -0.27 & 0.70\\
& & & & & &\\
\hline
\end{tabular}
\end{center}
\end{table}

The calculated vacancy formation energy (0.34 eV) is essentially
less than for Nb$_{1-x}$B$_2$ (E$_{vf} \sim$ 1.1 eV\cite{Shein3})
where the metal vacancies were found
experimentally\cite{Yamamoto,Takeya,Takeya1}. Hence, their
presence is very probably in AgB$_2$ and the stoichiometry of
AgB$_2$ based materials will be very sensitive to the synthesis conditions.\\
The redistribution of the charge states near V$^{Ag}$ is
demonstrated in Fig. 2. As is seen, in the Ag-deficient boride
the B-B bonds remain almost undistorted. No new bonds going
through the vacancy are formed. The difference map shows a weak
density accumulation along the B$\longrightarrow$V$^{Ag}$
direction, however the absence of the electron localization on
the vacancy clearly follows from $\Delta\rho$ contour
distribution in the planar (100) silver sheet, i.e the charge
states of silver vacancies are close to neutral. In the vicinity
of vacancy only insignificant deformation of $\Delta\rho$
contours appears along the direction Ag$\longrightarrow$V$^{Ag}$.
Figure 3 shows the changes in the DOS near the Fermi level with
Ag-vacancies concentration in Ag$_{1-x}$B$_2$. The Fermi level
slightly shifts towards the low energies with increasing x, while
the dependence of N(E$_F$) is not sensitive to vacancy
concentration. The N(E$_F$) values for Ag$_{0.875}$B$_2$ and
Ag$_{0.750}$B$_2$ are 0.88 and 0.92 states/eVcell,
respectively.\\
In summary, we presented the results of band structure
calculations for silver-deficient AgB$_2$ performed by the VASP
method. It was established that vacancies are likely to appear in
AgB$_2$: the energy of metal vacancies formation in silver
diboride is the least among all 4d metal diborides. Thus, the
nonstoiciometry on Ag-sublattice will be easily achievable, i.e.
for AgB$_2$ it is possible to expect the wide homogeneity region.
Our analysis indicates that metal vacancies in AgB$_2$ do not
lead to essential reduction of N(E$_F$) which seems to be
critical in depressing of the superconductivity in this system.

Acknowledgment.\\

This work was supported by the RFBR, grants 02-03-32971 and
04-03-32082.


\begin{figure*}[!htb]
\vskip  0cm
\begin{tabular}{c}

\includegraphics[width=12.0 cm,clip]{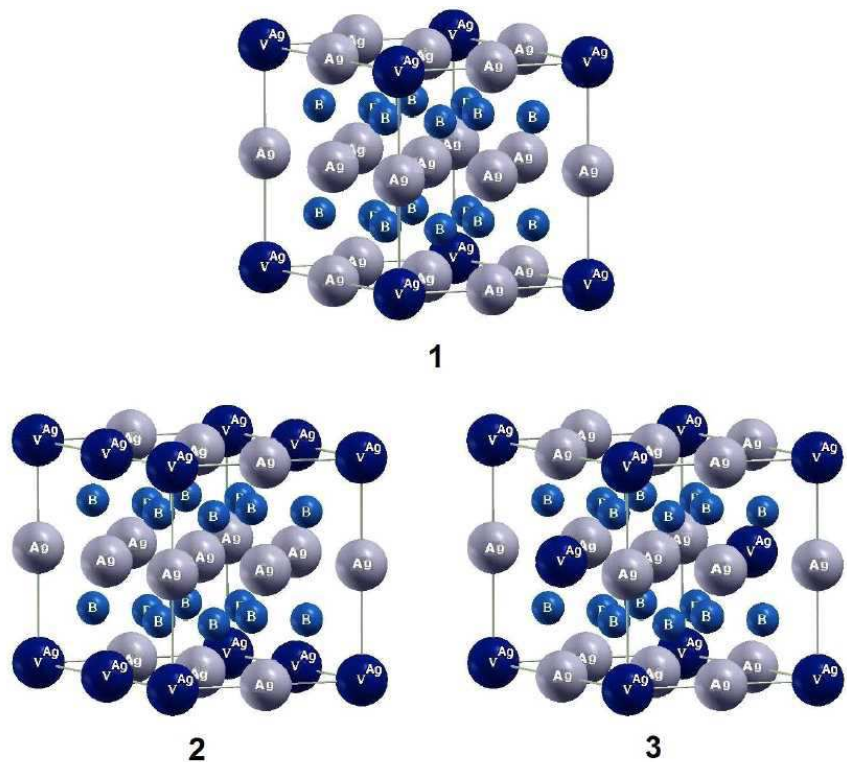} \\

\end{tabular}
\vspace{-0.02cm} \caption[a] { \small Fig. 1.Structural models
for silver-deficient Ag$_{1-x}$B$_{2}$: 1 - Ag$_{0,875}$B$_{2}$
and 2,3 - Ag$_{0.750}$B$_{2}$. The possible distribution of Ag
vacancies (V$^{Ag}$) are presented: 2 - the alternation of
complete and V$^{Ag}$ - containing sheets (along z-axis, model I)
and 3 - the "uniform" V$^{Ag}$ distribution (model II),
respectively.}
\end{figure*}

\begin{figure*}[!htb]
\vskip  0cm
\begin{tabular}{c}

\includegraphics[width=8.0 cm,clip]{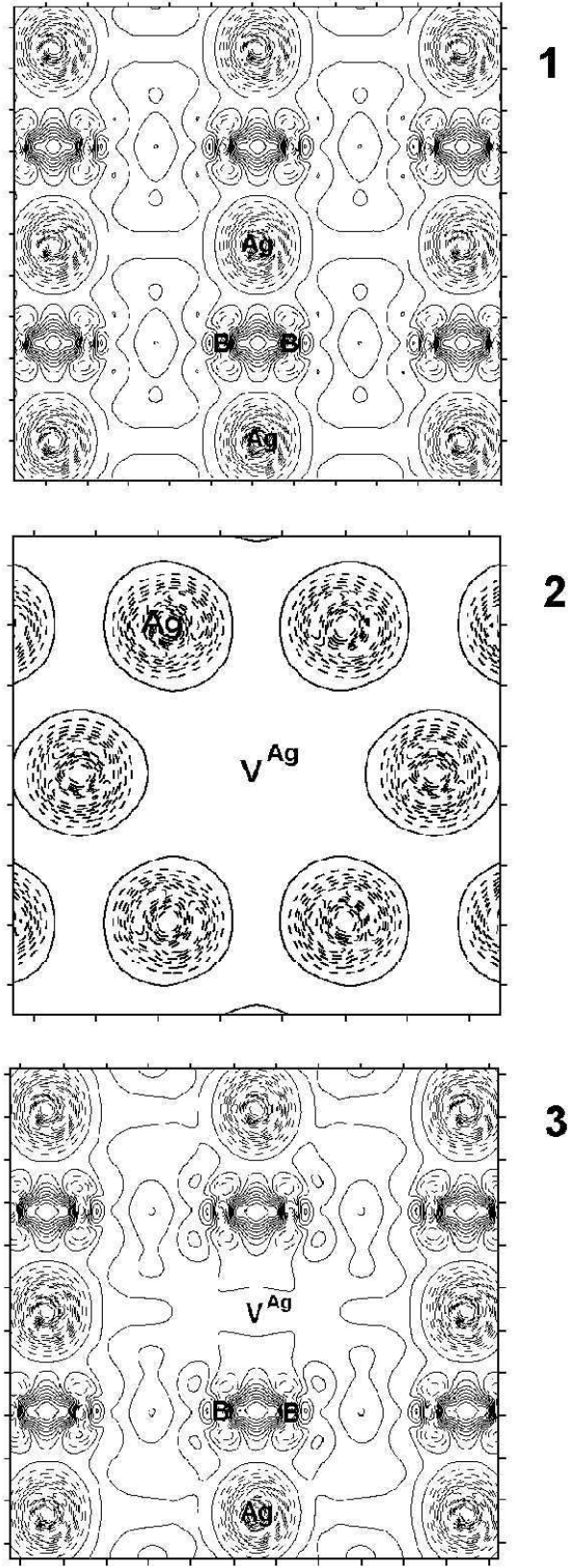}\\

\end{tabular}

\vspace{-0.02cm} \caption[a] { \small Fig. 2. Inter-plane
charge-density difference ($\Delta\rho$) maps for AgB$_2$ (1),
Ag$_{0.875}$B$_{2}$ (3) and $\Delta\rho$ map in (100) silver
sheet of Ag$_{0.875}$B$_2$. In the $\Delta\rho$ plots, solid and
dashed lines indicate an increase and a decrease of the electron
density $\Delta\rho^{AgB_2}$ relative to the atomic ($\rho^{Ag}$,
$\rho^B$) superposition, respectively.}
\end{figure*}

\begin{figure*}[!htb]
\vskip  0cm
\begin{tabular}{c}

\includegraphics[width=14.0 cm,clip]{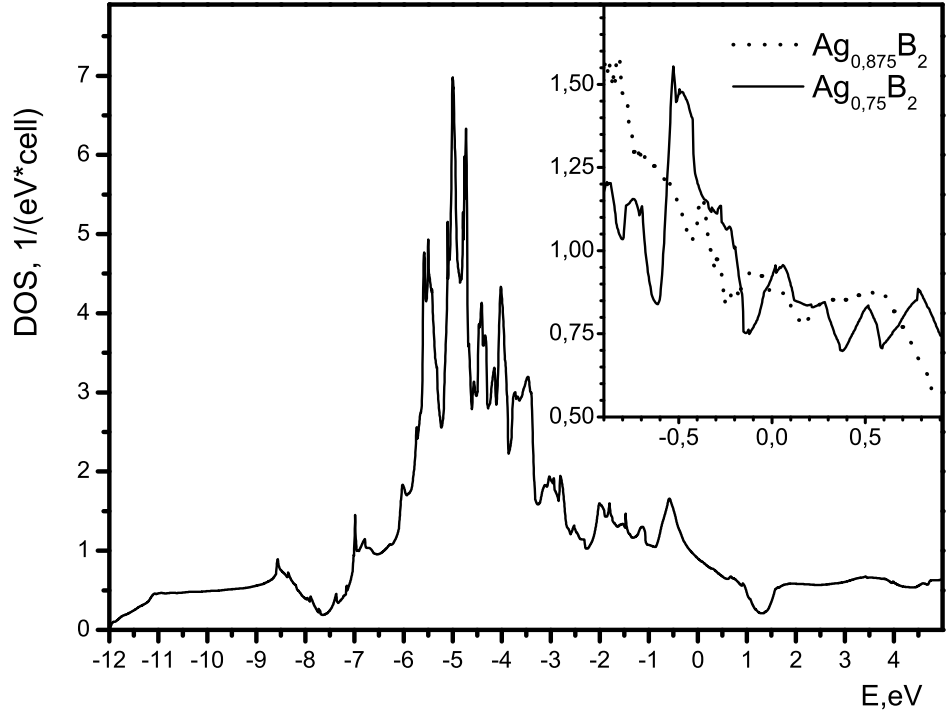}\\

\end{tabular}

\vspace{-0.02cm} \caption[a] { \small Density of states for
AgB$_2$. \textit{Inset} shows near-Fermi DOSs for
Ag$_{0.875}$B$_2$ and Ag$_{0.750}$B$_2$. E$_F$ = 0 eV.}
\end{figure*}

\end{document}